\documentstyle[prd,aps,preprint,epsfig]{revtex}
\tighten
\newcommand{\ind}[2]{^{#1}_{\text{\scriptsize #2}}}
\newcommand{\n}{{^{\text{\tiny N}}}\!}
\newcommand{\al}[2]{\alpha\ind{#1}{#2}}
\newcommand{\tal}[2]{\widetilde{\alpha}\ind{#1}{#2}}
\newcommand{\hal}[2]{\widehat{\alpha}\ind{#1}{#2}}
\newcommand{\tro}[2]{\widetilde{\rho}\ind{#1}{#2}}
\def\LQCD{$\Lambda_{\text{\scriptsize QCD}}$ }
\begin{document}
\draft
\title{New analytic running coupling in \\
       spacelike and timelike regions}
\author{A.\ V.\ Nesterenko\thanks{Electronic address:
nesterav@thsun1.jinr.ru}}
\address{Department of Physics, Moscow State University,\\
Vorobjovy Gory, Moscow, 119899, Russia}
\date{September 1, 2001}
\maketitle
\begin{abstract}
The new model for the QCD analytic running coupling, proposed
recently, is extended to the timelike region. This running coupling
naturally arises under unification of the analytic approach to QCD
and the renormalization group (RG) formalism. A new method for
determining the coefficients of the ``analytized'' RG equation is
elaborated. It enables one to take into account the higher loop
contributions to the new analytic running coupling (NARC) in a
consistent way. The expression for the new analytic running coupling,
independent of the normalization point, is obtained by invoking the
asymptotic freedom condition. It is shown that the difference between
the values of the NARC in respective spacelike and timelike regions
is rather valuable for intermediate energies. This is essential for
the correct extracting of the running coupling from experimental
data. The new analytic running coupling is applied to the description
of the inclusive $\tau$ lepton decay. The consistent estimation of
the parameter \LQCD is obtained here.
\end{abstract}

\section{Introduction}
     The description of hadron dynamics in the infrared (IR) region
remains an urgent problem of contemporary elementary particle theory.
The asymptotic freedom in quantum chromodynamics (QCD) enables one to
apply the standard perturbative approach at large momenta
transferred. However, a number of phenomena (for example, quark
confinement, nonvanishing vacuum expectation values, etc.) are beyond
such calculations. Moreover, the use of perturbation theory results
in known unphysical singularities in the IR region (for example, the
ghost pole appears in the expression for the running coupling).

     Recently a new model for the QCD analytic running coupling has
been proposed~\cite{PRD}. The basic idea of this model is unification
of the analytic approach to QCD~\cite{SolSh} with the renormalization
group (RG) formalism for recovering proper analytic properties of the
theory. An obvious advantage of the new analytic running coupling
(NARC) is that it incorporates IR enhancement with asymptotic freedom
behavior in a single expression and does not contain additional
parameters. In Refs.~\cite{PRD,ConfIV,ICMP2000} it was shown
explicitly that the new analytic running coupling, {\em without
invoking any additional assumptions}, leads to the quark-antiquark
potential rising at large distances. In addition, a reasonable
estimation of the parameter \LQCD was obtained there. The absence of
unphysical singularities in the physical region of
positive\footnote{In this paper a metric with the signature
$(-1,1,1,1)$ is used, so that $q^2>0$ corresponds to a spacelike
momentum transfer.} $q^2$ is also an appealing feature of NARC.

     The objective of this paper is to construct a consistent
continuation of the new analytic running coupling to the timelike
region and to elaborate the method for determination of the
coefficients of ``analytized'' RG equation (i.e., RG equation with
recovered proper analytic properties, see Sec.~II). The latter
enables one to take consistently into account the higher loop
contributions to the new analytic running coupling in the spacelike
region.

     The layout of the paper is as follows. In Sec.~II the new
analytic running coupling is considered in the spacelike region. The
method for determining the coefficients of the analytized RG equation
is proposed that enables one to take consistently into account the
higher loop contributions to NARC. By invoking the asymptotic freedom
condition the expression for the new analytic running coupling,
independent of the normalization point, is obtained. In Sec.~III the
continuation of the new analytic running coupling to the timelike
region is constructed. It is shown that distinction between the
corresponding values of NARC in the spacelike and timelike regions
becomes valuable in the intermediate energy region. This fact may be
important when extracting the QCD running coupling from experimental
data. Further, it is shown explicitly that for the timelike NARC the
respective $\beta$ function is proportional to the spectral density,
thus confirming the hypothesis due to Schwinger. In Sec.~IV the
results on studies of the new model for the QCD analytic running
coupling are briefly summarized. In the framework of the approach
developed the description of the inclusive $\tau$ lepton decay is
performed. The obtained value of the parameter \LQCD fairly well
agrees with its previous estimations. This implies the applicability
of the new analytic running coupling to description of the both
typical perturbative and intrinsically nonperturbative processes of
quantum chromodynamics. In the Conclusion (Sec.~V) the obtained
results are formulated in a compact way, and further studies in this
approach are outlined.

\section{New analytic running coupling in the spacelike region}
     In paper~\cite{PRD} a new model for the QCD analytic running
coupling has been proposed. The model is based on the unification of
the so-called analytic approach to QCD~\cite{SolSh} with the RG
formalism. Obvious advantage of the new analytic running coupling is
incorporation, in a single expression, of IR enhancement\footnote{It
is worth noting here that such a behavior of invariant charge follows
from the analysis of the Schwinger-Dyson equations~\cite{AlekArbu}.}
and asymptotic freedom. Remarkably, the model contains no additional
parameters. Similarly to the standard case, \LQCD remains the only
characterizing parameter of the theory. By making use of the NARC in
Refs.~\cite{PRD,ConfIV,ICMP2000} the confining interquark potential
was explicitly derived without invoking any additional assumptions,
the acceptable estimation of the parameter \LQCD being obtained. It
is also to be noted that NARC is free of unphysical singularities in
the physical region $q^2>0$.

     According to Ref.~\cite{PRD}, in the spacelike (Euclidean)
region $q^2>0$ the new analytic running coupling at the $\ell$-loop
level, $\n\al{(\ell)}{an}(q^2)$, is defined as the solution of the
analytized RG equation
\begin{equation}
\label{NARCEqn}
\frac{d\,\ln\bigl[\n\tal{(\ell)}{an}(q^2)\bigr]}{d\,\ln q^2} =
- \left\{\sum_{j=0}^{\ell-1}
B_j \Bigl[\tal{(\ell)}{s}(q^2)\Bigr]^{j+1}
\right\}_{\text{$\!$an}}.
\end{equation}
Here $\tal{(\ell)}{s}(q^2)$ is the $\ell$-loop perturbative running
coupling, $\widetilde{\alpha}(q^2)\equiv\alpha(q^2)\,\beta_0/(4\pi)$.
The curly brackets $\left\{S(q^2)\right\}_{\text{an}}$ mean the
``analytization'' (i.e., the recovering of proper analytic properties
in the $q^2$ variable) of the function $S(q^2)$ by making use of the
K\"all\'en-Lehmann spectral representation
\begin{equation}
\Bigl\{S(q^2)\Bigr\}_{\text{$\!$an}} \equiv \int_{0}^{\infty}
\!\frac{\varrho(\sigma)}{\sigma+q^2}\, d\sigma,
\end{equation}
where the spectral density $\varrho(\sigma)$ is determined by the
initial (perturbative) expression for $S(q^2)$:
\begin{equation}
\varrho(\sigma) = \frac{1}{2\pi i}\,
\lim_{\varepsilon \to 0_{+}}
\Bigl[S(-\sigma-i\varepsilon)-S(-\sigma+i\varepsilon)\Bigr].
\end{equation}
In addition, according to the model proposed in~\cite{PRD} the
solution of the RG equation~(\ref{NARCEqn}) before analytization must
be the perturbative running coupling at the respective loop level. In
other words, at any loop level the relationship
\begin{equation}
\label{BjDefEqn}
\frac{d\,\ln\bigl[\tal{(\ell)}{s}(q^2)\bigr]}{d\,\ln q^2} = -
\sum_{j=0}^{\ell-1} B_j \Bigl[\tal{(\ell)}{s}(q^2)\Bigr]^{j+1}
\end{equation}
holds. It is this condition that enables one to determine
consistently the coefficients $B_j$ on the right-hand side of
Eq.~(\ref{NARCEqn}).

     For this purpose let us consider the standard RG equation for
the invariant charge $g(\mu)$ (see, e.g., Refs.~\cite{BgSh,Ynd})
\begin{equation}
\label{StdRgEqn}
\frac{1}{g(\mu)}\,\frac{d\,g(\mu)}{d\,\ln\mu} =
\beta\Bigl(g(\mu)\Bigr).
\end{equation}
For the $\beta$ function the perturbative expansion
\begin{equation}
\beta\Bigl(g(\mu)\Bigr) = - \left\{
\beta_{0}\left[\frac{g^2(\mu)}{16 \pi^2}\right] +
\beta_{1}\left[\frac{g^2(\mu)}{16 \pi^2}\right]^2 + \ldots \right\}
\end{equation}
takes place, where $\beta_{0} = 11 - 2 \,n_{\text f}/3$, $\,\beta_{1}
= 102 - 38 \,n_{\text f}/3$, and $n_{\text f}$ is the number of
active quarks. Further, Eq.~(\ref{StdRgEqn}) can be rewritten in the
following form:
\begin{equation}
\label{NewPertEqn}
\frac{d\,\ln \bigl[g^2(\mu)\bigr]}{d\,\ln\mu^2} =
\beta\Bigl(g(\mu)\Bigr).
\end{equation}
Introducing the standard notations $\al{}{s}(\mu^2)=g^2(\mu)/(4\pi)$
and $\tal{}{s}(\mu^2)=\al{}{s}(\mu^2)\,\beta_{0}/(4\pi)$,
Eq.~(\ref{NewPertEqn}) at the $\ell$-loop level can be reduced to
Eq.~(\ref{BjDefEqn})
\begin{equation}
\label{EqnBjDef}
\frac{d\,\ln\bigl[\tal{(\ell)}{s}(\mu^2)\bigr]}{d\,\ln \mu^2} = -
\sum_{j=0}^{\ell-1} \beta_j
\left[\frac{\tal{(\ell)}{s}(\mu^2)}{\beta_{0}}\right]^{j+1},
\end{equation}
if one puts $B_j=\beta_j/(\beta_0)^{j+1}$. Actually, in this case
Eq.~(\ref{BjDefEqn}) is nothing but the standard perturbative RG
equation that defines the invariant charge.

\begin{figure}
\noindent
\centerline{\epsfig{file=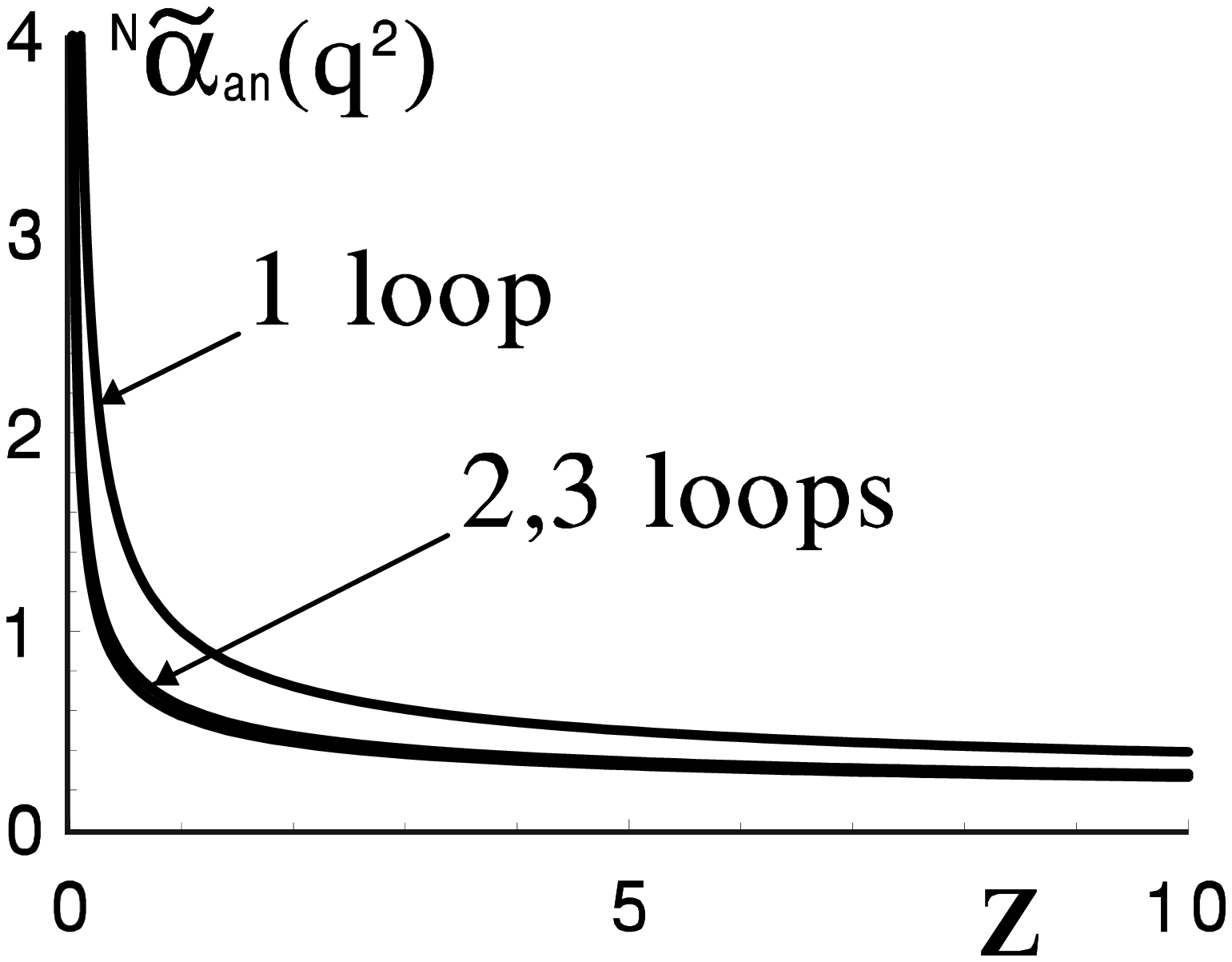, width=125mm}}
\caption{The new analytic running coupling
$^{\text{N}}\widetilde{\alpha}_{\text{an}}(q^2)$
(Eq.~(\protect\ref{NARCLLDef})) in the spacelike region at the one-,
two-, and three-loop levels, $z=q^2/\Lambda^2$.}
\label{NARCPlot}
\end{figure}

     Thus, in the spacelike region the new analytic running coupling
at the $\ell$-loop level $\n\al{(\ell)}{an}(q^2)$ is defined as the
solution of the equation
\begin{equation}
\label{NARCEqnDef}
\frac{d\,\ln\bigl[\n\tal{(\ell)}{an}(q^2)\bigr]}{d\,\ln q^2} =
- \left\{ \sum_{j=0}^{\ell-1}
B_j \Bigl[\tal{(\ell)}{s}(q^2)\Bigr]^{j+1}
\right\}_{\text{$\!$an}}, \quad
B_j = \frac{\beta_{j}}{\beta_{0}^{j+1}}.
\end{equation}
At the one-loop level Eq.~(\ref{NARCEqnDef}) can be integrated
explicitly with the result~\cite{PRD}
\begin{equation}
\label{NARC1LDef}
\n\al{(1)}{an}(q^2) =
\frac{4\pi}{\beta_0}\,\frac{z-1}{z\,\ln z}, \qquad
z=\frac{q^2}{\Lambda^2}.
\end{equation}
At the higher loop levels there is only the integral representation
for the new analytic running coupling. One should note from the very
beginning, that the solution of Eq.~(\ref{NARCEqnDef}) is determined
up to a constant factor due to the logarithmic derivative on its
left-hand side. In our previous studies this problem has been
eliminated by normalization of the solution to Eq.~(\ref{NARCEqnDef})
on its value at a point $q_0^2$ (see, e.g., Eq.~(2) in
Ref.~\cite{MPLA}). In this paper we propose to remove this ambiguity
in a more physical way. Indeed, this can be easily achieved by
invoking the condition of the asymptotic freedom, namely
$\n\al{(\ell)}{an}(q^2) \to \al{(\ell)}{s}(q^2)$ when $q^2 \to
\infty$ (in fact, it has already been employed at the one-loop level,
see Eq.~(\ref{NARC1LDef})).  Then, involving the similar condition
$\n\al{(\ell)}{an}(q^2) \to \n\al{(1)}{an}(q^2)$ when $q^2 \to
\infty$, we obtain the following integral representation for the
$\ell$-loop new analytic running coupling (see Ref.~\cite{NARCMATHII}
for the details):
\begin{equation}
\label{NARCLLDef}
\n\al{(\ell)}{an}(q^2) = \frac{4\pi}{\beta_{0}} \,
\frac{z-1}{z\,\ln z}\,
\exp\!\left[\int_{0}^{\infty}\! \Delta {\cal R}^{(\ell)}(\sigma)
\,\ln\!\left(1 + \frac{\sigma}{z}\right)\, \frac{d \sigma}{\sigma}
\right],
\end{equation}
where $\Delta {\cal R}^{(\ell)}(\sigma) =
{\cal R}^{(\ell)}(\sigma) - {\cal R}^{(1)}(\sigma)$, and
\begin{eqnarray}
\label{SpDensDef}
{\cal R}^{(\ell)}(\sigma) &=& \frac{1}{2 \pi i}
\lim_{\varepsilon \to 0_{+}}
\sum_{j=0}^{\ell-1}B_j
\biggl\{\Bigl[\tal{(\ell)}{s}(-\sigma-i\varepsilon)\Bigr]^{j+1}
\nonumber\\
&& -\Bigl[\tal{(\ell)}{s}(-\sigma+i\varepsilon)\Bigr]^{j+1}\biggr\},
\quad \sigma \ge 0.
\end{eqnarray}
It is worth noting here that for the NARC the K\"all\'en-Lehmann
representation holds
\begin{equation}
\label{NARCIntRepSL}
\n\al{(\ell)}{an}(q^2) = \frac{4\pi}{\beta_{0}}\int_{0}^{\infty}
\frac{\n\tro{(\ell)}{}(\sigma)}{\sigma+z}\,d\sigma, \quad
z=\frac{q^2}{\Lambda^2},
\end{equation}
where the spectral density at the one-loop level is
\begin{equation}
\label{NewRho1LDef}
\n\tro{(1)}{}(\sigma) =
\left(1+\frac{1}{\sigma}\right)\frac{1}{\ln^2\!\sigma+\pi^2},
\end{equation}
and the $\ell$-loop spectral density $\n\tro{(\ell)}{}(\sigma)$ is
defined below (see Eq.~(\ref{SpecDensDef})). Figure~\ref{NARCPlot}
shows the new analytic running coupling~(\ref{NARCLLDef}) computed at
the one-, two-, and three-loop levels. It is clear from these curves
that NARC possesses a quite good loop stability (for the properties
of the new analytic running coupling see
Refs.~\cite{MPLA,NARCMATHII}).

     An important merit of the new analytic running coupling is that
it allows one to obtain the confining quark-antiquark ($q\bar q$)
potential without invoking any additional
assumptions~\cite{PRD,ConfIV}. Besides, the interquark potential
contains no adjustable parameters (it depends solely on the distance
$r$ and the parameter $\Lambda_{\text{\scriptsize QCD}}$), and at
small distances it has the standard behavior prescribed by the
asymptotic freedom.  Comparison of the $q\bar q$ potential generated
by the NARC~\cite{PRD} with the phenomenological (Cornell) potential
and with the lattice simulation data~\cite{Bali}, as well as the
estimation of the gluon condensate value on the base of NARC give a
consistent value\footnote{This estimation corresponds to the one-loop
level with three active quarks.} of the parameter \LQCD:
$\Lambda\simeq 600 \pm 90$~MeV. This can be considered as an evidence
of the fact that the new analytic running coupling adequately takes
into account the nonperturbative nature of quantum chromodynamics
(see also Ref.~\cite{NPQCD01}).

\section{New analytic running coupling in the timelike region}
     In the previous section the definition of the new analytic
running coupling in the spacelike region has been given. However, for
consistent description of a number of QCD processes (for example,
$\tau$ lepton decay or $e^{+}e^{-}$ annihilation to hadrons) one has
to use the continuation of the running coupling to the timelike
region.

     In Ref.~\cite{MiltSol} the procedure of continuation of the
invariant charge from the spacelike region to the timelike region
(and vise versa) was elaborated by making use of the dispersion
relation for the Adler $D$ function. In particular, if the running
coupling in the spacelike region is $\alpha(q^2)$, then its
consistent continuation to the timelike region is defined by the
integral
\begin{equation}
\label{ARCTLDef}
\widehat{\alpha}(s) = \frac{1}{2 \pi i}
\int_{s+i\varepsilon}^{s-i\varepsilon}
\frac{d\,\zeta}{\zeta}\,\alpha(-\zeta),
\quad s=-q^2>0,
\end{equation}
where the integration contour goes from the point $s+i\varepsilon$ to
the point $s-i\varepsilon$ and lies in the region of analyticity of
the function $\alpha(-\zeta)$. Here and further the running coupling
in the spacelike region is denoted by $\alpha(q^2)$ and in the
timelike region by $\widehat{\alpha}(s)$.

     In order to simplify Eq.~(\ref{ARCTLDef}) let us choose the
integration contour in the following way. From the point
$s+i\varepsilon$ the integration path goes in a parallel way with the
real axis to infinity, then along the circle with an infinitely large
radius it goes counter-clockwise to the point $(\infty - i
\varepsilon)$, and then it goes in a parallel way with the real axis
to the point $s - i \varepsilon$. As a result, the continuation of
the one-loop new analytic running coupling $\n\al{(1)}{an}(q^2)$ to
the timelike region is given by
\begin{equation}
\label{NARCIntRepTL}
\n\hal{(1)}{an}(s) = \frac{4\pi}{\beta_{0}}\,
\int_{s/\Lambda^2}^{\infty}~\!\!
\n\tro{(1)}{}(\zeta)\,\frac{d\zeta}{\zeta}
\end{equation}
with the spectral density $\n\tro{(1)}{}(\zeta)$ defined in
Eq.~(\ref{NewRho1LDef}).

     The account of higher loop corrections leads to significant
technical complications. So, at the $\ell$-loop level for the new
analytic running coupling in the timelike region the integral
representation of the same form takes place:
\begin{equation}
\label{NARCTLDef}
\n\hal{(\ell)}{an}(s) = \frac{4 \pi}{\beta_0} \,
\int_{s/\Lambda^2}^{\infty}
\n\tro{(\ell)}{}(\zeta) \, \frac{d \zeta}{\zeta},
\end{equation}
where
\begin{eqnarray}
\label{SpecDensDef}
\n\tro{(\ell)}{}(\zeta) &=& \n\tro{(1)}{}(\zeta) \, \exp\!\left[
\int_{0}^{\infty} \Delta {\cal R}^{(\ell)}(\sigma)
\, \ln \left|1 - \frac{\sigma}{\zeta} \right| \,
\frac{d \sigma}{\sigma} \right] \nonumber\\
&& \times \left[
\cos \psi^{(\ell)}(\zeta) +
\frac{\ln \zeta}{\pi} \sin \psi^{(\ell)}(\zeta)
\right],
\end{eqnarray}
${\cal R}^{(\ell)}(\sigma)$ is defined in Eq.~(\ref{SpDensDef}), and
\begin{equation}
\psi^{(\ell)}(\zeta) = \pi \int_{\zeta}^{\infty}
\Delta {\cal R}^{(\ell)}(\sigma)\, \frac{d \sigma}{\sigma}.
\end{equation}
In Eq.~(\ref{SpecDensDef}) the principle value of the integral is
assumed.

     It is interesting to note here that, in our approach, a simple
relation holds between the $\beta$ function, corresponding to the
running coupling in the timelike region (Eq.~(\ref{NARCTLDef})), and
the relevant spectral density:
\begin{equation}
\beta\Bigl(\widehat{\alpha}(s)\Bigr) \equiv
\frac{d\,\left[\n\hal{(\ell)}{an}(s)\right]}{d\, \left[\ln s\right]}
=-\frac{4\pi}{\beta_{0}} \, \n\tro{(\ell)}{}(s).
\end{equation}
Thus the $\beta$ function in question proves to be proportional to
the spectral density $\n\tro{(\ell)}{}(s)$. Obviously, this result
completely agrees with the attempts, originated by
Schwinger~\cite{Schwinger}, to find a direct physical interpretation
of the $\beta$ function (the so-called Schwinger hypothesis, see
Refs.~\cite{MiltSol,Milton}).

\begin{figure}
\noindent
\centerline{\epsfig{file=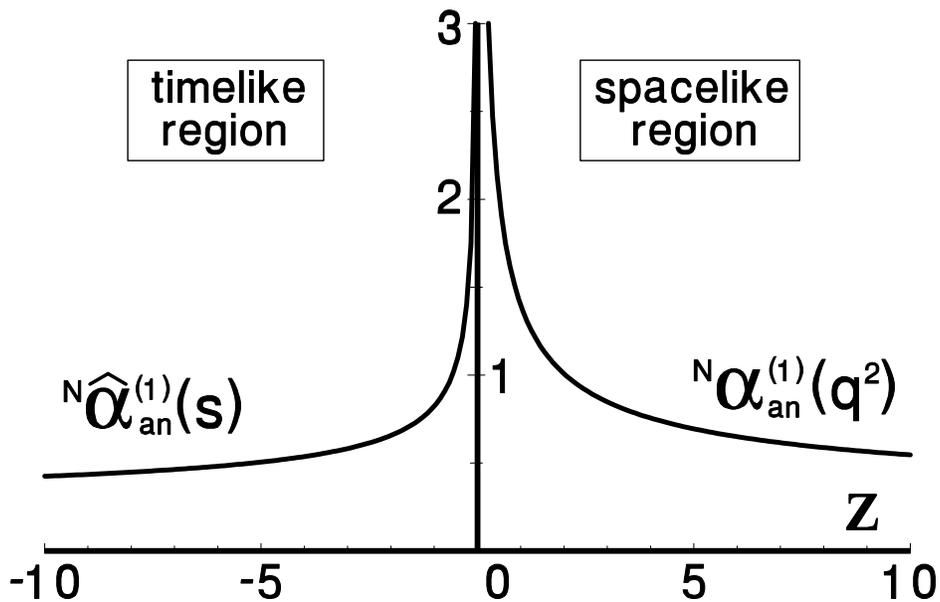, width=125mm}}
\caption{The one-loop new analytic running coupling in the spacelike
($q^2>0$) and timelike ($s=-q^2>0$) regions
(Eqs.~(\protect\ref{NARC1LDef}) and (\protect\ref{NARCIntRepTL}),
respectively), $z=q^2/\Lambda^2$.}
\label{TSLPlot}
\end{figure}

\begin{figure}
\noindent
\centerline{\epsfig{file=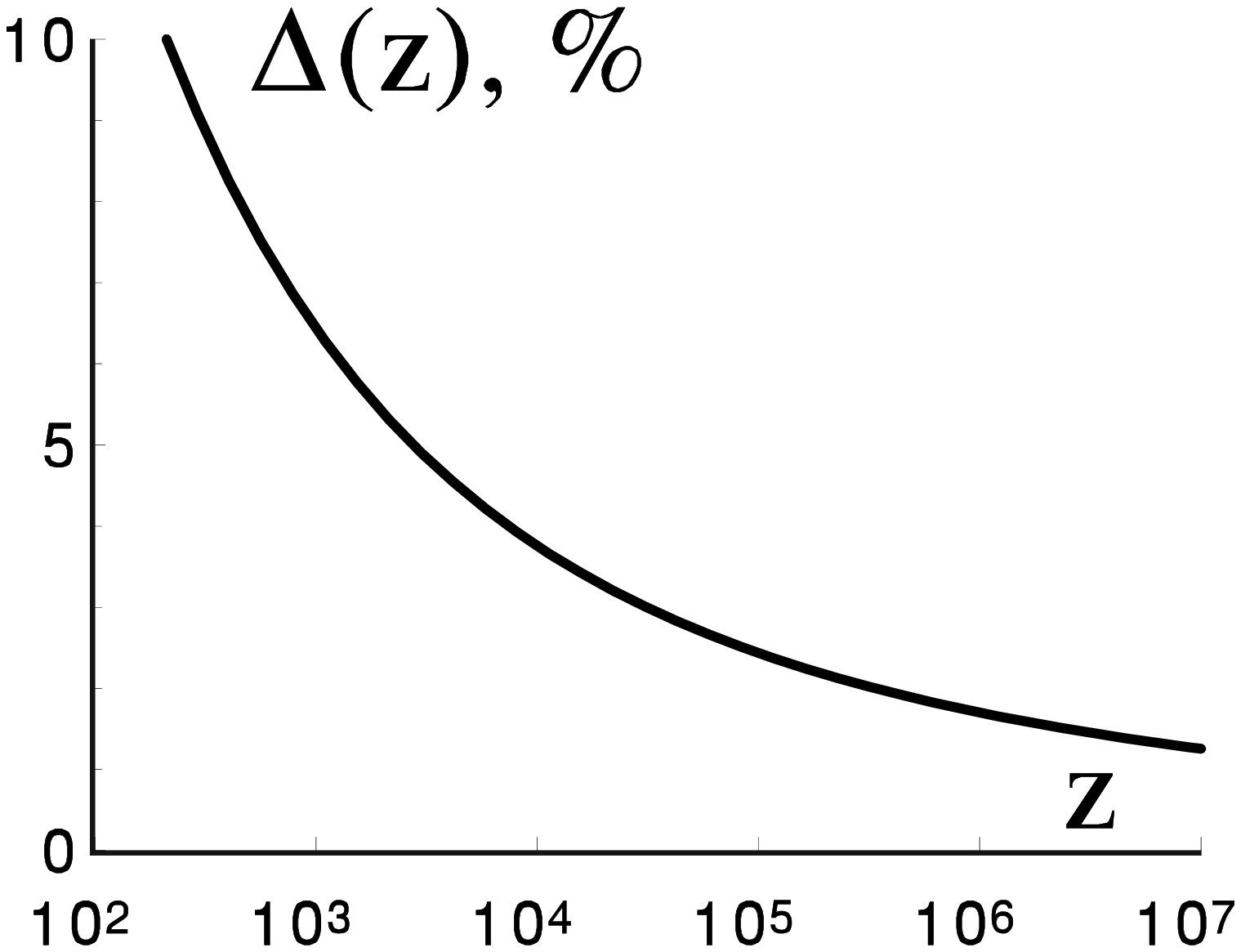, width=125mm}}
\caption{The relative difference
$\Delta(z)=\left[~\!^{\text{N}}\alpha_{\text{an}}
^{\text{(1)}}(q^2)/~\!
^{\text{N}}\widehat{\alpha}_{\text{an}}
^{\text{(1)}}(-q^2)-1\right]\times
100 \%$ between the values of the one-loop new analytic running
coupling in the spacelike and timelike regions, $z=q^2/\Lambda^2$.}
\label{TSLRatioPlot}
\end{figure}

     The plots of the functions $\n\al{(1)}{an}(q^2)$ and
$\n\hal{(1)}{an}(s)$ are shown in Fig.~\ref{TSLPlot}. At large
values of the arguments these expressions have identical behavior
prescribed by the asymptotic freedom $\alpha(q^2) \sim 1/\ln |z|$.
However, for intermediate energies the difference between these
couplings is rather valuable. The ratio of the one-loop new analytic
running coupling in the spacelike region $\n\al{(1)}{an}(q^2)$ to its
continuation to the timelike region $\n\hal{(1)}{an}(-q^2)$ is
presented in Fig.~\ref{TSLRatioPlot}. It follows from this figure
that the deviation increases when approaching the IR region, and, in
particular, rises up to about $10 \%$ when $\sqrt{s} \simeq 10$~GeV.
Apparently, this circumstance should be taken into account when
extracting the QCD running coupling from experimental data.

\section{Discussion}
     The basic idea of the analytic approach to quantum field theory
can be treated as an attempt to recover explicitly the proper
analytic properties in the $q^2$ variable for the relevant physical
quantities. A concrete realization of this approach may be different.
For example, in the original model by Shirkov and
Solovtsov~\cite{SolSh} the analytization procedure is applied to the
perturbative invariant charge. Unlike this proposal, our new model
for the QCD analytic running coupling~\cite{PRD} seeks to recover the
proper analytic properties of the RG equation. More precisely, the
new analytic running coupling is derived as the solution of the RG
equation with the $\beta$ function ``improved'' by the analytization
procedure (see also Refs.~\cite{PRD,NPQCD01} for the details). From
the general point of view, in this case one can anticipate new
properties of the invariant charge. Indeed, in our model there is the
IR enhancement of the new analytic running coupling, while the
invariant charge due to Shirkov and Solovtsov tends to a finite value
in the infrared limit (the IR freezing of the running coupling). It
is this property of our model that ultimately leads to the confining
quark-antiquark potential. Further, the behavior of the NARC in the
UV region is completely determined by the asymptotic freedom in QCD.
Thus, in our model~\cite{PRD} the IR enhancement and the UV
asymptotic freedom are involved in a single expression for the new
analytic running coupling without introducing any additional
parameters into the theory.  It is important to emphasize here that
such a behavior of the invariant charge is in agreement with the
Schwinger-Dyson equations (see Ref.~\cite{AlekArbu}).

     Let us turn now to the perturbative phenomena. One of the QCD
processes, the most sensitive to the low energy behavior of the
invariant charge, is the inclusive semileptonic branching ratio of
the $\tau$ lepton decay, $R_{\tau}$. Therefore, we address the
consideration of this process restricting ourselves, for simplicity,
to the one loop level.

     We proceed from the nonstrange part of the $R_{\tau}$ ratio
associated with the vector quark currents
\begin{equation}
\label{RTauDef}
R_{\tau,\text{\tiny V}} = \frac{N_{\text{c}}}{2}\, |V_{\text{ud}}|^2
S_{\text{\tiny EW}} \, \left(1 + \delta_{\text{\tiny QCD}}\right),
\end{equation}
where $N_{\text{c}} = 3$ is the number of quark colors,
$|V_{\text{ud}}|=0.9735 \pm 0.0008$ denotes the
Cabibbo-Kobayashi-Maskawa matrix element~\cite{EPJC2000},
$S_{\text{\tiny EW}}=1.0194 \pm 0.0040$ is the electroweak
factor~\cite{MarSi}, and $\delta_{\text{\tiny QCD}}$ is the QCD
correction (see, e.g., Refs.~\cite{MilSolSol,GeIo} and references
therein). A recent experimental measurement of the
ratio~(\ref{RTauDef}) by ALEPH Collaboration gave~\cite{ALEPH}
$R_{\tau,\text{\tiny V}} = 1.775 \pm 0.017$.

     In accordance with the standard prescription the one-loop QCD
correction in our approach is determined by the integral
\begin{equation}
\label{DeltaQCDDef}
\delta_{\text{\tiny QCD}} = \frac{2}{\pi}\int_{4m^2}^{M_{\tau}^2}
\frac{d\, s}{M_{\tau}^2} \left(1 - \frac{s}{M_{\tau}^2}\right)^2
\left(1 + 2 \frac{s}{M_{\tau}^2}\right) \n\hal{(1)}{an}(s),
\end{equation}
where $\n\hal{(1)}{an}(s)$ is the new analytic running coupling in
the timelike region~(\ref{NARCIntRepTL}), and $m$ is the light quark
mass, with its world average value $m = (4.25 \pm
1.75)$~MeV~\cite{EPJC2000}. It is worth noting here that there is no
need to involve the contour integration in Eq.~(\ref{DeltaQCDDef}),
since NARC~(\ref{NARCIntRepTL}) does not contain unphysical
singularities in the region $s > 0$. In other words, the integration
in Eq.~(\ref{DeltaQCDDef}) can be performed in a straightforward way.
By introducing the notations $x=s/M_{\tau}^2$, $x_0=4m^2/M_{\tau}^2$,
and $c_0 = - (2x_0-2x_0^3+x_0^4)$, one can represent
Eq.~(\ref{DeltaQCDDef}) in a form convenient for integration
\begin{eqnarray}
\delta_{\text{\tiny QCD}} &=&
\frac{4\pi}{\beta_0}\int_{x_0}^{1}
\left(2x-2x^3+x^4+c_0\right) \n\tro{(1)}{}\!
\left(x\frac{M_{\tau}^2}{\Lambda^2}\right)\,\frac{d\,x}{x}
\nonumber \\
&& + \, \frac{4\pi}{\beta_0}\,(1+c_0)
\int_{M_{\tau}^2/\Lambda^2}^{\infty}\n\tro{(1)}{}(\sigma)\,
\frac{d\,\sigma}{\sigma},
\end{eqnarray}
where the spectral density $\n\tro{(1)}{}(\sigma)$ is defined in
Eq.~(\ref{NewRho1LDef}).

     For the value of $R_{\tau,\text{\tiny V}}$ given
above~\cite{ALEPH} we obtain the estimation $\Lambda = (560 \pm
70)$~MeV for two active quarks (the uncertainty accounts the errors
in the values of $R_{\tau,\text{\tiny V}}$, $|V_{\text{ud}}|$,
$S_{\text{\tiny EW}}$, and $m$). However, in order to compare this
result with earlier estimations, one should continue it to the region
of three active quarks. This gives the value $\Lambda = (517 \pm
70)$~MeV that perfectly agrees with the previous estimations. We
remind that this value of the parameter \LQCD corresponds to the
one-loop level with three active quarks.

     Thus, in the framework of the model under consideration one
succeeds in obtaining a consistent, with its previous estimations,
value of the parameter \LQCD by making use of the experimental data
on the inclusive $\tau$ lepton decay, the world average value of
light quark mass being employed.

     From all this we may infer that the new model for the QCD
analytic running coupling substantially incorporates, in a consistent
way, both perturbative (high energy) and nonperturbative (low energy)
behavior of quantum chromodynamics.

\section{Conclusion}
     This paper is a further development of the new model for the QCD
analytic running coupling proposed in~\cite{PRD}. This running
coupling naturally arises under unification of the analytic approach
to QCD with the RG formalism. The elaborated method for determination
of the coefficients of the analytized RG equation enables one to take
consistently into account the higher loop contributions to the new
analytic running coupling. The invoking of the asymptotic freedom
condition allows one to derive the expression for the NARC
independent of the normalization point. The continuation of the new
analytic running coupling to the timelike region is constructed. It
is demonstrated that the difference between the respective values of
the new analytic running coupling in the spacelike and timelike
regions is considerable for intermediate energies. This fact seems to
be important for extracting the running coupling from experimental
data. It is shown that for the timelike new analytic running coupling
the hypothesis due to Schwinger is confirmed. In the framework of the
new model for the QCD analytic running coupling the description of
the inclusive $\tau$ lepton decay is performed. The obtained value of
the parameter \LQCD is in a good agreement with its previous
estimations.

     Thus, the application of the new analytic running coupling to
the description of various physical phenomena (both, perturbative and
intrinsically nonperturbative ones) gives a consistent estimation of
the parameter $\Lambda_{\text{\scriptsize QCD}}$. This fact, as well
as the agreement of the IR behavior of the NARC with Schwinger-Dyson
equations, and ability to obtain, without invoking any additional
assumptions, the confining quark-antiquark potential, implies
that the new model for the QCD analytic running coupling
substantially incorporates, in a consistent way, both perturbative
(high energy) and nonperturbative (low energy) behavior of quantum
chromodynamics.

     In further studies it would be useful to obtain simple explicit
expressions which approximate the new analytic running coupling in
the spacelike and timelike regions with account of the higher loop
corrections.

\acknowledgments
     The author is grateful to Dr.\ I.\ L.\ Solovtsov for valuable
discussions and useful comments and to Professor G.\ S.\ Bali for
supplying the relevant data on the lattice calculations. The partial
support of RFBR (Grant No.\ 00-15-96691) is appreciated.

\vspace{5mm}

\vspace{5mm}

\begin{references}
\bibitem{PRD} A.\ V.\ Nesterenko, Phys.\ Rev.\ D {\bf 62}, 094028
         (2000); hep-ph/9912351.
\bibitem{SolSh} D.\ V.\ Shirkov and I.\ L.\ Solovtsov, Phys.\ Rev.\
         Lett.\ {\bf 79}, 1209 (1997); hep-ph/9704333; Theor.\ Math.\
         Phys.\ {\bf 120}, 482 (1999); hep-ph/9909305.
\bibitem{ConfIV} A.\ V.\ Nesterenko, in {\it Proceedings of the 4th
         International Conference on Quark Confinement and the Hadron
         Spectrum, Vienna, Austria, 2000}, edited by W.\ Lucha
         (HEPHY, Vienna, in press), hep-ph/0010257.
\bibitem{ICMP2000} A.\ V.\ Nesterenko, in {\it Book of Abstracts of
         the XIII International Congress on Mathematical Physics,
         London, United Kingdom, 2000}, p.\ 97.
\bibitem{AlekArbu} A.\ I.\ Alekseev and B.\ A.\ Arbuzov, Mod.\ Phys.\
         Lett.\ A {\bf 13}, 1747 (1998); hep-ph/9704228.
\bibitem{BgSh} N.\ N.\ Bogolyubov and D.\ V.\ Shirkov, {\it
         Introduction to the Theory of Quantized Fields}
         (Interscience, New York, 1980).
\bibitem{Ynd} F.\ J.\ Yndurain, {\it Quantum Chromodynamics}
         (Springer-Verlag, Berlin, Heidelberg, 1983).
\bibitem{MPLA} A.\ V.\ Nesterenko, Mod.\ Phys.\ Lett.\ A {\bf 15},
         2401 (2000); hep-ph/0102203.
\bibitem{NARCMATHII} A.\ V.\ Nesterenko (in preparation).
\bibitem{Bali} G.\ S.\ Bali {\it et al.}, Phys.\ Rev.\ D {\bf 62},
         054503 (2000); hep-lat/0003012.
\bibitem{NPQCD01} A.\ V.\ Nesterenko, in {\it Proceedings of the
         Sixth Workshop on Non-Perturbative Quantum Chromodynamics,
         Paris, France, 2001} (to be published); hep-ph/0106305.
\bibitem{MiltSol} K.\ A.\ Milton and I.\ L.\ Solovtsov, Phys.\ Rev.\
         D {\bf 55}, 5295 (1997); hep-ph/9611438.
\bibitem{Schwinger} J.\ Schwinger, Proc.\ Natl.\ Acad.\ Sci.\ U.S.A.\
         {\bf 71}, 3024 (1974); {\bf 71}, 5047 (1974).
\bibitem{Milton} K.\ A.\ Milton, Phys.\ Rev.\ D {\bf 10}, 4247
         (1974).
\bibitem{EPJC2000} Particle Data Group, D.\ E.\ Groom {\it et al.},
         Eur.\ Phys.\ J.\ C {\bf 15}, 1 (2000).
\bibitem{MarSi} W.\ J.\ Marciano and A.\ Sirlin, Phys.\ Rev.\ Lett.\
         {\bf 56}, 22 (1986); {\bf 61}, 1815 (1988).
\bibitem{MilSolSol} K.\ A.\ Milton, I.\ L.\ Solovtsov, and O.\ P.\
         Solovtsova, Phys.\ Rev.\ D {\bf 64}, 016005 (2001);
         hep-ph/0102254.
\bibitem{GeIo} B.\ V.\ Geshkenbein, B.\ L.\ Ioffe, and K.\ N.\
         Zyablik, hep-ph/0104048.
\bibitem{ALEPH} ALEPH Collaboration, R.\ Barate {\it et al.}, Eur.\
         Phys.\ J.\ C {\bf 4}, 409 (1998).
\end{references}
\end{document}